\documentclass[sigconf,nonacm]{acmart}

\AtBeginDocument{%
  \providecommand\BibTeX{{%
    Bib\TeX}}}

\usepackage{multirow}
\usepackage{booktabs}
\usepackage[linesnumbered,ruled,vlined]{algorithm2e}
\usepackage[normalem]{ulem}
\usepackage{balance}
\usepackage{threeparttable}
\usepackage{url}

\usepackage{graphicx}
\usepackage{textcomp}
\usepackage{xcolor}
\usepackage{hyperref}
\usepackage{subcaption}
\usepackage{caption}

\def\BibTeX{{\rm B\kern-.05em{\sc i\kern-.025em b}\kern-.08em
    T\kern-.1667em\lower.7ex\hbox{E}\kern-.125emX}}

\renewcommand\footnotetextcopyrightpermission[1]{}
\begin{document}

\title{Rethinking Clause Management for CDCL SAT Solvers}

\author{Yalun Cai}
\authornote{These authors contributed equally to this work.}
\affiliation{%
  \institution{The Chinese University of Hong Kong}
  \city{Hong Kong}
  \country{China}
}
\email{ylcai25@cse.cuhk.edu.hk}

\author{Xindi Zhang}
\authornotemark[1] 
\authornote{Corresponding author.}
\affiliation{%
  \institution{Institute of Software, Chinese Academy of Sciences}
  \city{Beijing}
  \country{China}
}
\email{zhangxd@ios.ac.cn}

\author{Zhengyuan Shi}
\affiliation{%
  \institution{The Chinese University of Hong Kong}
  \city{Hong Kong}
  \country{China}
}
\email{zyshi21@cse.cuhk.edu.hk}

\author{Mengxia Tao}
\affiliation{%
  \institution{National Center of Technology Innovation for EDA}
  \city{Nanjing}
  \country{China}
}
\email{taomengxia@nctieda.com}

\author{Qiang Xu}
\authornotemark[2] 
\affiliation{%
  \institution{The Chinese University of Hong Kong}
  \city{Hong Kong}
  \country{China}
}
\email{qxu@cse.cuhk.edu.hk}

\begin{abstract}

Boolean Satisfiability (SAT) solving underpins a wide range of applications in Electronic Design Automation (EDA), particularly formal verification.
However, this paper observes that the mainstream clause reduction heuristic in modern SAT solvers becomes ineffective in the critical domain of complex arithmetic circuit verification, such as multipliers. 
On these instances, the dominant Literal Block Distance (LBD) metric for measuring clause quality degrades into a simple value of clause length, without any perception of dynamic clause usage during solving.
To address this issue, a novel clause reduction mechanism is proposed, which is entirely independent of LBD. Its core idea is to decouple and handle separately the two most fundamental characteristics of learnt clauses—inherent lineage and dynamic usage patterns—thereby avoiding the efficiency degradation caused by inappropriately mixing these properties. Experiments show that our method consistently improves mainstream solvers and achieves speedups of up to 5.74× on complex arithmetic circuit problems, while maintaining comparable performance on general-purpose benchmarks. These results challenge the prevailing LBD-centric clause quality metric for clause management.

\end{abstract}

\maketitle


\section{Introduction} \label{Sec:Intro}

\begin{figure*}[t]
    \centering
    \begin{subfigure}[b]{0.4\textwidth}
        \centering
        \includegraphics[width=\textwidth]{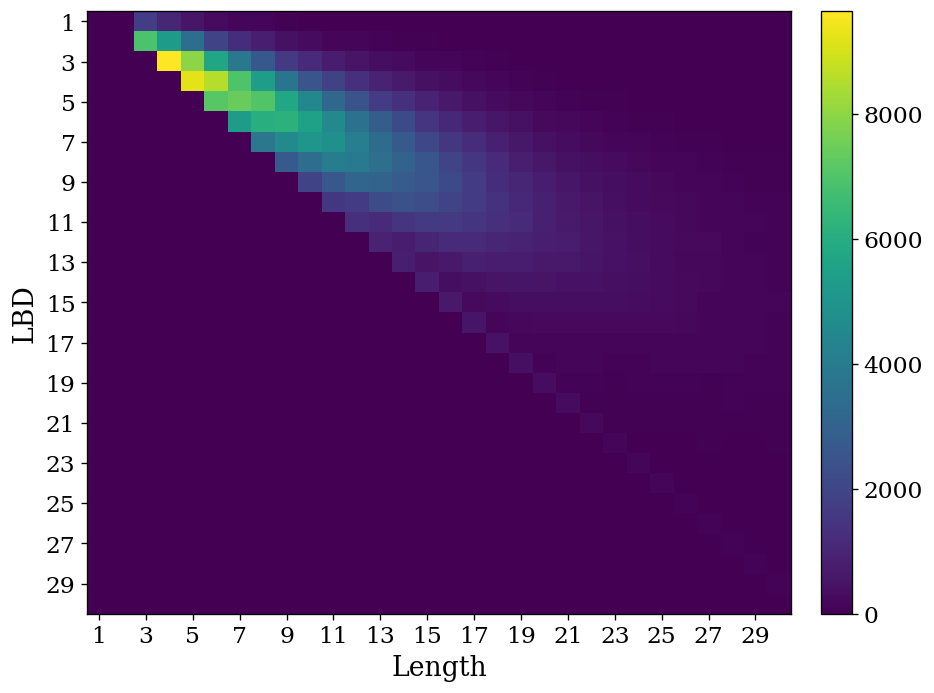}
        \caption{General Circuit Equivalence Checking Instance}
        \label{fig:sub1}
    \end{subfigure}
    \hfill
    \begin{subfigure}[b]{0.4\textwidth}
        \centering
        \includegraphics[width=\textwidth]{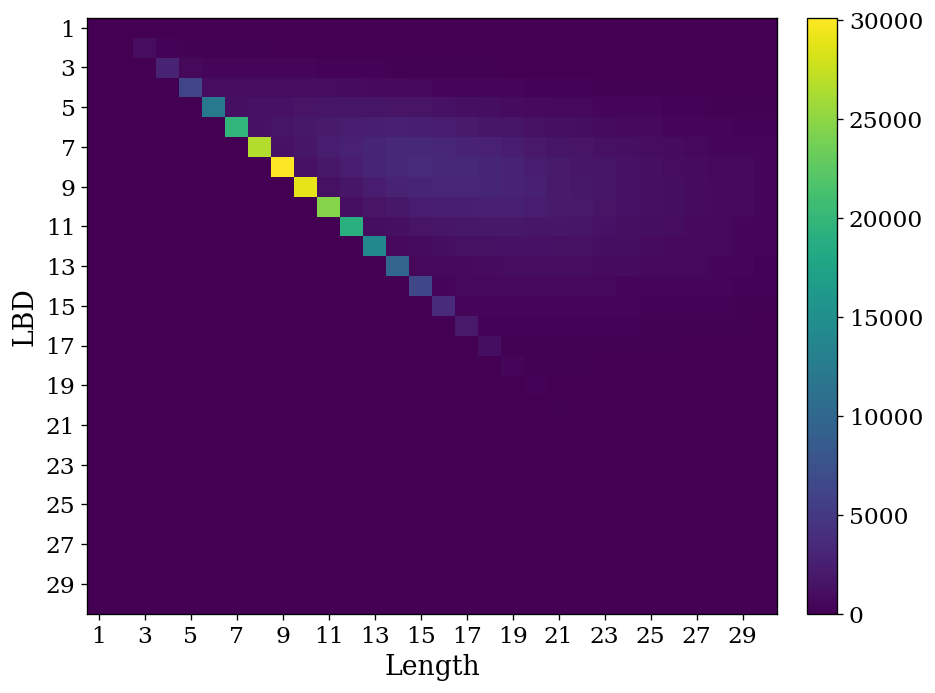}
        \caption{Arithmetic Circuit Equivalence Checking Instance}
        \label{fig:sub2}
    \end{subfigure}
    \caption{Heatmap distributions of learnt clauses, correlating clause length (x-axis) with LBD (y-axis). For (a) LBD provides a rich and dynamic quality metric that is distinct from static clause length. For (b), the distribution collapses into a sharp diagonal line, indicating the LBD of a learnt clause is almost always consistent with its length. Therefore, LBD-based clause reduction heuristic fails to consider the dynamic behavior of variable decision beyond the static clause length already offers.}
    \label{fig:motivation}
    \vspace{-10pt}
\end{figure*}

The Boolean Satisfiability (SAT) problem, which determines whether a given Boolean formula has a satisfying assignment, is a canonical NP-complete problem~\cite{cook2023complexity}. 
The SAT problem has become a cornerstone of automated reasoning and plays a central role in numerous applications, particularly in Electronic Design Automation (EDA), as well as in artificial intelligence and planning.

To handle the ever-growing scale and complexity of SAT instances in practical domains, modern SAT solvers, such as kissat, CaDiCaL, and MiniSat~\cite{biere2024cadical, sorensson2005minisat, minisatpaper}, have evolved for decades and incorporated efficient solving algorithms. Despite their success on standard benchmarks, mainstream SAT solvers with general heuristics exhibit limitations when applied to EDA tasks, particularly in the complex verification of arithmetic circuits. These circuits, such as multipliers and dividers, while small in scale, present a high level of solving complexity. 

Such limitations motivate a deeper investigation into the core engine of these solvers, namely the Conflict-Driven Clause Learning (CDCL) framework~\cite{marques2021conflict}. After encoding the given problem instance into Conjunctive Normal Form (CNF), consisting of multiple clauses, CDCL solvers iteratively explore the search space and learn new clauses from conflicts to prune infeasible assignments. 
Effective management of learnt clauses is crucial to the performance of CDCL solvers, with the central factor being the assessment of learnt clause quality, which dictates the removal of uninformative or low-utility clauses.
For over fifteen years, the Literal Block Distance (LBD)~\cite{audemard2009glucose} metric has been the dominant measure. LBD, defined as the number of distinct decision levels in a clause, theoretically combines a clause's static length\footnote{Clause length refers to the number of literals contained in a clause.} with its dynamic usage patterns during decision propagation. 

However, we observe that for most cases of arithmetic circuit equivalence checking, the LBD metric has nearly degraded into a simple clause length measure. As highlighted in Figure~\ref{fig:motivation}, this degradation indicates that standard clause reduction algorithms fail to organically integrate static clause features and dynamic decision behaviors in this domain. Therefore, we raise a critical question: \textit{Is it still appropriate to use the current LBD-based clause management algorithm for these complex verification tasks?}

To answer this question, we propose a novel two-stage learnt clause management algorithm. The core idea lies in completely decoupling and separately handling the two most fundamental characteristics of learnt clauses — \textit{dynamic usage patterns} and \textit{inherent lineage} — thereby avoiding the efficiency degradation caused by the inappropriate mixing of these two essential properties. 
In the first stage, we focus exclusively on the actual usage patterns of learnt clauses, maximizing the retention of clauses that have demonstrated superior performance through the two mechanisms by which clauses can primarily influence solver behavior: Boolean Constraint Propagation (BCP) and conflict analysis. 
In the second stage, for clauses that have not yet demonstrated their potential, we rely solely on clause lineage (i.e., clause length) to break ties, giving promising but underutilized clauses an opportunity to prove themselves and minimizing the risk of erroneous deletion. 

Specifically, the main contributions of our work include:

\begin{itemize}
    \item  We challenge the supremacy of LBD, the dominant clause quality metric that has held sway for over fifteen years, particularly in the domain of complex arithmetic circuit verification problems.
    \item We propose completely decoupling clause lineage from actual usage patterns, and design a novel clause reduction mechanism that is entirely independent of LBD.
    \item The effectiveness of our method on complex datapath arithmetic units, represented by multipliers, through detailed experiments, and conduct comprehensive data analysis.
\end{itemize}

We integrate our algorithm into mainstream SAT solvers and validate its effectiveness on both complex arithmetic verification and general-purpose benchmarks. Experimental results demonstrate that our method consistently improves solver performance on complex arithmetic verification problems, reducing kissat's PAR2 score by approximately 40\,\% and nearly doubling the number of instances successfully solved by MiniSat. 
On instances that neither timeout before nor after integrating our method, our approach achieves up to 5.74× speedup and an average speedup of 3.88× on MiniSat. Furthermore, on the 2022 SAT Competition benchmarks, our method enables kissat to solve 4 additional instances with a 4.7\,\% PAR-2 score reduction, demonstrating significant improvement on arithmetic verification problems without compromising performance on other domains.

\section{Related Work} \label{Sec:Related}




\subsection{SAT and Arithmetic Verification}

The Boolean Satisfiability (SAT) problem asks whether there exists an assignment that satisfies a given Boolean formula. Modern SAT solvers predominantly rely on the Conflict-Driven Clause Learning (CDCL) algorithm~\cite{marques2021conflict}. The CDCL algorithm operates through an iterative process of unit propagation, conflict analysis, and backtracking, with Boolean Constraint Propagation (BCP) serving as the primary inference engine that drives the core deduction mechanism. CDCL solvers employ CNF (Conjunctive Normal Form) as the standard representation, where a formula is structured as a conjunction of clauses $\phi = (C_1 \land C_2 \land \ldots)$, with each clause being a disjunction of literals (a literal is either a Boolean variable or its negation), e.g., $C_i = (x_1 \lor \neg x_2 \lor \ldots)$. SAT is NP-complete, making it computationally challenging for large instances, while remaining fundamental in hardware and software verification, planning, scheduling, and combinatorial design~\cite{biere2009handbook}.

Verifying the functional equivalence of two circuits is of fundamental importance in EDA~\cite{qian2025xsat}. However, equivalence checking for complex arithmetic circuits poses significant computational challenges, with multipliers representing one of the most difficult classes of circuits to verify. To address these challenges, SAT-based reasoning provides an effective approach for such verification tasks.


\subsection{Clause Management of Modern SAT Solvers}
A primary reason for the effectiveness of CDCL solvers is their ability to learn clauses during conflict analysis, enabling effective search space pruning. However, due to memory constraints and the computational overhead of BCP, CDCL solvers must periodically eliminate relatively less useful learnt clauses to maintain solving efficiency. Thus, clause management constitutes a critical mechanism for ensuring the overall efficiency of CDCL solvers.

SOTA CDCL solvers, exemplified by the kissat series, employ a three-tier clause reduction mechanism based on LBD, following the insights from \cite{oh2015between}. Learnt clauses are partitioned into tiers by LBD, with each tier applying different retention policies based on recent conflict analysis usage. Highest-tier clauses (smallest LBD) are preserved if they have been used within a long time window; middle-tier clauses require recent usage; while lowest-tier clauses (largest LBD) are always filtered. The filtered clauses from all tiers are then sorted by LBD (with length as a tie-breaker), and those with the largest LBD and length are deleted. This mechanism has two limitations. First, it depends on LBD's quality characterization, which relies on LBD's ability to organically integrate clause lineage with usage patterns. When this integration fails for certain problems, efficiency degrades. Second, the redundant use of LBD in both filtering and sorting creates poor coordination, causing low-LBD but low-quality clauses to be preserved while high-LBD but critical clauses are removed.

Before the prevalence of the aforementioned SOTA solvers, the MiniSat solver\cite{minisatpaper} was highly valued by both academia and industry for its clear, concise code structure and lightweight implementation. The core idea of MiniSat's clause reduction mechanism transfers VSIDS, a representative branching heuristic, to clause management. It maintains an activity score for each learnt clause using an "additive increase, multiplicative decrease" pattern: activity increases when a clause participates in conflict analysis, and all clause activities are uniformly decreased multiplicatively by a decay factor after each conflict. When reduction is needed, clauses with lower activity scores are prioritized for deletion. However, MiniSat's reduction mechanism has three significant limitations. First, clause management differs fundamentally from variable selection because the set of candidate variables remains fixed while learnt clauses are continuously added, which makes the direct transfer of branching heuristics inappropriate. Second, it ignores a clause's lineage (i.e., its length), rendering quality assessment incomplete and preventing integration of a clause's inherent lineage with its dynamic usage patterns. Third, while relying entirely on dynamic usage for quality assessment, MiniSat's approach is still incomplete: learnt clauses affect solver behavior through both conflict analysis and BCP, yet the mechanism ignores the latter.

\section{Methodology} \label{Sec:Method}

\subsection{Overview of the Proposed Algorithm}


From the preceding discussion, LBD is the prevalent clause quality metric that combines reasoning depth (via decision levels) and clause length, effectively guiding clause management in many contexts. 
However, when applied to complex arithmetic circuit problems, the lack of meaningful decision-level structure causes LBD to degenerate into a proxy for clause length alone (Fig.~\ref{fig:motivation}). 

To address this limitation, we propose a novel LBD-free, two-stage clause management scheme that evaluates clause quality by fully decoupling a clause’s inherent lineage from its dynamic usage patterns, thereby avoiding the efficiency losses incurred when the two are conflated into a single heuristic.

Algorithm~\ref{alg:nolbd-core} outlines the proposed method. 
Stage 1 focuses exclusively on dynamic usage, retaining clauses that excel on that dimension, whereas Stage 2 considers only inherent lineage, giving clauses that were quiet in Stage 1 but still appear promising a chance to survive. By isolating one fundamental characteristic per stage, the procedure eliminates the need for LBD and prevents the inefficiencies that occur when LBD fails to integrate lineage and usage on this benchmark family.

\subsection{Stage 1: Dynamic Usage Filtering}
Stage 1 focuses exclusively on the dynamic usage frequency of learnt clauses. Throughout the entire solver execution, learnt clauses influence the solver’s behavior through, and only through, the following two mechanisms:

\begin{itemize}
    \item  \textbf{Boolean Constraint Propagation(BCP)}: when a clause becomes unit under the current partial assignment, it forces the remaining literal to be assigned, directly steering the search and pruning the current branch.
    \item \textbf{Conflict Analysis}: when a clause serves as a reason during conflict resolution, it shapes the learnt clause and the ensuing backtrack level, thereby redirecting the search trajectory.
\end{itemize}

Thus, evaluating clause quality by its functional contribution through these two channels is more principled and equitable than relying exclusively on its inherent pedigree; hence, Stage 1 prioritizes actual usage over lineage.
In our approach, we assess the quality of a learnt clause by the frequency with which it influences the solver through two channels. Specifically, we count how many times the clause actively affects the search processes. 

To track this, we assign a score to each learnt clause. The score is initialized to 1 when the clause is created (lines 1--2). This reflects its origins in conflict analysis, which immediately reduces the search space and guides the solver. Afterwards, the score increases by 1 each time the clause is used in BCP or in a subsequent conflict analysis (lines 3--4).



Relying solely on a clause’s cumulative usage frequency has a key limitation because it fails to reflect how the search region evolves. 
As newly learned clauses interact and create new pruning regions, they may gradually overshadow the pruning effect provided by older clauses, even if those clauses were once essential. To mitigate this issue, we introduce an aging mechanism. At every fixed interval of \textit{T} conflicts, the score of each clause with a positive value is reduced by one (lines 5--7). This process allow removing clauses whose pruning power has been superseded and ensures that only those that actively influence the current search continue to receive high scores.

When a reduction is triggered, Stage 1 retains every learnt clause whose score remains positive and forwards only the zero-score clauses to Stage 2. This allows truly impactful clauses to stay in the database, while clauses that have stopped contributing—and now only consume resources—are passed along for further screening.

\subsection{Stage 2: Lineage-Oriented Selection}



Stage 2 receives the clauses that Stage 1 filtered out and further prunes them using only their inherent lineage—namely, clause length, the sole universally accepted clause-quality metric aside from LBD~\cite{liang2018mlrestart}.

Our motivation for Stage 2 arises from the observation that Stage 1 may inadvertently filter out several categories of valuable clauses. First, newly learned clauses often have considerable potential but lack sufficient opportunities to demonstrate their effectiveness because of their recent introduction. These clauses clearly merit additional chances before being discarded.

Second, since a large number of clauses are removed at once, some clauses that previously appeared ineffective may become essential after other clauses that covered the same pruning region have been eliminated. It is common for several overlapping clauses to constrain the same region of the search space. Because the pruning effect is distributed across these clauses, none of them appears strongly used, which makes them all seem weak in Stage 1. As a result, Stage 1 may incorrectly remove all of them. Therefore, Stage 2 should ensure that at least one representative clause is preserved for every repeatedly covered pruning region.


Third, a CDCL solver explores different parts of the problem instance over time. 
Clauses that are relevant to the region currently under search tend to be used frequently, while those associated with other regions may be overlooked for long periods. When the solver later moves to a previously unvisited or recently revisited region, these overlooked clauses may again become important. Stage 2 preserves space for such high-potential clauses that appear underused simply because their associated region has not been encountered recently.

In summary, Stage 2 provides a second opportunity for the three types of clauses identified above.


Since Stage 1 already makes full use of the dynamic usage information of clauses, Stage 2 relies exclusively on clause length, which reflects their inherent lineage, to break ties. This approach avoids the inefficiency that could result from improperly combining these two distinct characteristics. In this stage, candidate learnt clauses are prioritized solely by length, and a configurable fraction of the longest clauses is removed.

\subsection{Reduction Triggering Policy and Amount}

After introducing the above mechanisms, the remaining question is when a reduction should be triggered and how many clauses should be reduced.

\paragraph{Reduction Triggering Policy}
In this work we simply adopt kissat’s strategy. Algorithm~\ref{alg:should-reduce} illustrates the decision procedure. In short, the solver schedules the gap between two consecutive reductions as $\sqrt{1}, \sqrt{2}, \sqrt{3}, \ldots$ times a base interval of $1000$ conflicts. We reuse this strategy because it adapts well to instances of varying difficulty. As long as the search keeps running without termination, the instance is presumably hard and will require more learnt clauses (or a longer refutation) before the solver returns a result. The slowly increasing sequence therefore allows the clause database to grow only at a moderate pace, matching the demand of harder instances without being overly aggressive.

\paragraph{Reduction Percentage}
When a reduction is triggered, 0-score clauses in Stage 1 are forwarded to Stage 2 for reduction.
To determine the removal fraction, we adopt a strategy inspired by kissat~\cite{biere2024cadical}, in which the rate is dynamically adapted. 

Specifically, the fraction interpolates between a lower bound of\,0.50 and an upper bound of\,0.90: starting near\,0.50 during early reductions, it asymptotically approaches\,0.90 as the number of reductions\,\(r\) increases, following the relation:

\[
\text{percent} = 0.90 - \frac{0.90 - 0.50}{\log_{10}(r + 9)}.
\]

This computed percentage then determines the exact proportion of candidates to be deleted from the list sorted by clause length.

\SetKw{KwDecay}{decay}
\SetKw{KwReduce}{reduce}
\SetKw{KwRemove}{remove}
\SetKwProg{Proc}{Procedure}{}{end}

\SetKwFunction{ProcLearn}{\textbf{OnLearnedClause}}
\SetKwFunction{ProcUse}{\textbf{OnClauseUse}}
\SetKwFunction{ProcDecay}{\textbf{OnPeriodicDecay}}
\SetKwFunction{ProcReduce}{\textbf{TwoStageReduction}}

\begin{algorithm}[t]
  \caption{Two-Stage Clause Management without LBD}
  \label{alg:nolbd-core}
  \KwIn{decay interval $\mathit{T}$}

  \Proc{\ProcLearn{$c$}}{
    $score(c) \gets 1$
  }

  \Proc{\ProcUse{$c$}}{
    $score(c) \gets score(c) + 1$
  }

  \Proc{\ProcDecay{}}{
    \ForEach{learnt clause $c$}{
      $score(c) \gets \max\{0, score(c) - 1\}$
    }
  }

  \Proc{\ProcReduce{}}{
    (Stage 1) $\mathcal{Z} \gets \{\, c \mid score(c) = 0 \,\}$\; 
    (Stage 2) Sort $\mathcal{Z}$ by clause length in descending order; 
    and select $\mathcal{C}\subset \mathcal{Z}$, the leading fraction of $\mathcal{Z}$\;
    \KwRemove{all deleted clauses $\mathcal{C}$ from the database}\;
  }

  \While{the solver is active}{
    \If{a clause is learned}{\ProcLearn{c}}
    \If{a clause participates in BCP or conflict analysis}{\ProcUse{c}}
    \If{the number of conflicts mod $\mathit{T}$ is 0}{\ProcDecay{}}
    \If{the reduction limit is hit}{\ProcReduce{}}
  }
\end{algorithm}

\SetKw{KwReturn}{return}
\SetKw{KwTrue}{true}
\SetKw{KwFalse}{false}
\SetKwProg{Fn}{Function}{}{end}

\begin{algorithm}[t]
  \caption{Predicate \textsc{ShouldReduce}}
  \label{alg:should-reduce}
  \KwData{%
    current conflict count $C$;\\
    number of reductions already performed $r$;\\
    next reduction threshold $L$ (initially $L \gets 1000\sqrt{1}$)}
  \KwResult{\KwTrue\ if a reduction should be triggered, else \KwFalse}

  \Fn{\textsc{ShouldReduce}$(C, r, L)$}{
    \If{$C < L$}{
      \KwReturn \KwFalse\;
    }
    $r \gets r + 1$,  $L \gets L + 1000 \sqrt{\,r+1\,}$\;
    \KwReturn \KwTrue\;
  }
\end{algorithm}
\section{Experiments} \label{Sec:Experiment}

\paragraph{Implementation Details}
Our method is implemented in both the kissat solver (version 4.0.2)~\cite{biere2024cadical} and the MiniSat solver~\cite{sorensson2005minisat}. 
kissat series represents the state-of-the-art, having dominated the main track of the SAT Competitions from 2020 to 2025, whereas MiniSat remains the most widely adopted in industry and is frequently employed as a baseline for performance benchmarking. All relevant source code is available in our GitHub repository\footnote{https://github.com/RethinkingClauseManagement/RethinkingClauseManagement}.

\paragraph{Datasets}
All competitors are evaluated on two datasets. 
The \textit{first dataset} probes their ability to handle complex arithmetic circuit problems and consists of 60 multiplier LEC instances: 33 originate from intrinsically challenging industrial cases converted into CNF format by ABC~\cite{mishchenko2007abc}, and 27 are drawn from the main track of recent SAT Competitions (15 from 2024~\cite{kondratiev2024benchmarks} and 12 from 2025~\cite{cai2025bit}). All 60 instances are unsatisfiable, mirroring the industrial reality that most verification workloads are UNSAT and that solver efficiency on such cases is far more consequential. Instances are available in our GitHub repository\footnote{https://github.com/RethinkingClauseManagement/RethinkingClauseManagement/\\tree/master/datasets/circuit}. 
The \textit{second dataset} is the main-track benchmark of the SAT Competition 2022 (SC22), adopted to evaluate solver performance on general-purpose instances; for fairness, we avoid using benchmarks from recent years, as some categories from recent competitions have already been included in the first dataset.

\paragraph{Setups and Metrics}
All experiments were conducted on a high-performance computing cluster equipped with 2 Intel(R) Xeon(R) Platinum 8153 @ 2.00GHz, totaling 128 physical cores, and 2TB RAM. The system operates on Ubuntu 22.04 LTS (64-bit). The gcc version is 11.4.0. 
For the first dataset, the time limit is set to 10,000 seconds to accommodate particularly hard instances, while for the second dataset we follow the SAT Competition convention and use a 5,000-second cutoff.

Some key metrics are reported:
The \textit{average PAR-2 Score}(PAR-2) is the average runtime, penalizing unsolved instances with twice the cutoff time.
Lower PAR-2 values indicate better performance.
\textit{\#Solved} refers to the number of instances that have been successfully solved; \textit{\#Better} denotes the number of instances in which the solver's performance is better than its competitor.


\paragraph{Experiments on Parameter Settings}
The sole hyperparameter of our approach is the decay interval \textit{\textbf{T}}. In this subsection, we evaluate its impact on both kissat and MiniSat. Because our method primarily targets complex arithmetic circuit problems, all experiments here are conducted on the first dataset. Table~\ref{tab:param-comparison} summarizes the respective results. They show that the optimal settings of \textit{T} for kissat and MiniSat are 4096 and 2048, respectively; therefore, we adopt these configurations for the remainder of the evaluation.

\begin{table}[tbp]
  \centering
  \caption{Effect of $T$ on kissat and MiniSat Performance}
  \label{tab:param-comparison}
  \scalebox{0.73}{
  \begin{tabular}{cccccc}
    \toprule
    \multicolumn{3}{c}{kissat} & \multicolumn{3}{c}{MiniSat} \\
    \cmidrule(r){1-3} \cmidrule(l){4-6}
    $T$ & \#Solved & PAR-2 & $T$ & \#Solved & PAR-2 \\
    \midrule
    1024 & 58 & 2166.13 & 1024 & 27 & 12492.21 \\
    2048 & 59 & 1796.21 & 2048 & \textbf{29} & \textbf{12177.66} \\
    4096 & \textbf{60} & \textbf{1724.01} & 4096 & 25 & 12886.96 \\
    8192 & \textbf{60} & 1725.38 & 8192 & 23 & 13330.16 \\
    \bottomrule
  \end{tabular}
  }
\end{table}



\subsection{Effectiveness Analysis}
\begin{table}[htbp]
\centering
\renewcommand{\arraystretch}{0.95}
\caption{Comparison on Complex Arithmetic Circuits}
\label{dataset1}
\scalebox{0.73}{
\begin{tabular}{lrr|rr|rr}
\hline
 &  &  & \multicolumn{2}{c|}{kissat} & \multicolumn{2}{c}{MiniSat} \\
Name & Vars & Clauses & origin & ours & origin & ours \\
\hline
KvW-12x11 & 5068 & 15159 & 6298.66 & \textbf{3476.53} & - & - \\
CvW-12x11 & 3111 & 9288 & 3181.44 & \textbf{1970.30} & - & - \\
DvK-11x10 & 4538 & 13573 & 733.68 & \textbf{601.31} & - & - \\
bwm\_dfo28 & 2967 & 9634 & 1420.57 & \textbf{1026.79} & - & - \\
CvD-11x11 & 2780 & 8297 & 794.83 & \textbf{777.08} & - & - \\
bwo\_dm28 & 2754 & 9111 & 1872.54 & \textbf{1280.20} & - & - \\
CvK-12x12 & 5462 & 16339 & - & \textbf{8202.60} & - & - \\
CvK-12x11 & 5253 & 15714 & 5416.89 & \textbf{3301.91} & - & - \\
KvW-11x10 & 4595 & 13744 & 729.47 & \textbf{606.83} & - & \textbf{9811.76} \\
bwo\_dfm29 & 2803 & 9267 & 698.89 & \textbf{507.19} & - & \textbf{8768.05} \\
bdm\_wm28 & 3124 & 10065 & 1688.22 & \textbf{1220.56} & - & - \\
CvW-12x12 & 3426 & 10231 & 5945.86 & \textbf{4476.83} & - & - \\
DvW-12x11 & 2861 & 8538 & 3962.77 & \textbf{2271.35} & - & - \\
bdo\_ado28 & 2540 & 8495 & 2702.40 & \textbf{1599.58} & - & - \\
bwm\_wo28 & 2964 & 9625 & 1530.09 & \textbf{1110.02} & - & - \\
bdm\_bwm & 3327 & 10653 & \textbf{1706.11} & 8653.96 & - & - \\
bdo\_adm28 & 2729 & 9022 & 2232.06 & \textbf{1583.00} & - & - \\
bdo\_ado29 & 2575 & 8611 & 935.78 & \textbf{689.35} & - & \textbf{9823.41} \\
wo\_dfmU27 & 2983 & 9722 & 2170.76 & \textbf{1652.67} & - & - \\
DvK-12x12 & 5189 & 15520 & - & \textbf{9593.52} & - & - \\
DvW-11x10 & 2357 & 7030 & 517.44 & \textbf{450.46} & - & \textbf{9583.85} \\
CvK-11x10 & 4745 & 14194 & 655.31 & \textbf{563.19} & - & \textbf{9445.01} \\
DvW-12x12 & 3153 & 9412 & 6552.19 & \textbf{5803.89} & - & - \\
KvW-10x10 & 3790 & 11331 & 272.48 & \textbf{266.32} & - & \textbf{3597.41} \\
bdm\_wo28 & 2948 & 9573 & 1748.21 & \textbf{1256.12} & - & - \\
CvK-11x11 & 5061 & 15140 & 2123.68 & \textbf{1507.14} & - & - \\
dfmU\_dm28 & 3166 & 10211 & 1066.36 & \textbf{866.05} & - & \textbf{9527.26} \\
m\_10\_t12 & 406 & 1627 & 46.83 & \textbf{43.31} & 1384.80 & \textbf{466.44} \\
m\_10\_t13 & 462 & 1866 & \textbf{38.84} & 41.96 & 1728.90 & \textbf{473.07} \\
m\_11\_t11 & 386 & 1488 & 94.62 & \textbf{83.27} & 3309.54 & \textbf{1156.10} \\
m\_11\_t12 & 452 & 1770 & 220.60 & \textbf{196.10} & 5776.35 & \textbf{2131.53} \\
m\_11\_t13 & 520 & 2054 & 306.40 & \textbf{285.17} & - & \textbf{2955.75} \\
m\_11\_t14 & 582 & 2324 & 283.98 & \textbf{252.94} & 8042.09 & \textbf{3073.63} \\
m\_11\_t15 & 641 & 2570 & 197.82 & \textbf{144.12} & 7243.98 & \textbf{1674.17} \\
m\_11\_t16 & 690 & 2787 & 105.68 & \textbf{88.27} & 3256.88 & \textbf{949.00} \\
m\_11\_t17 & 737 & 2988 & 50.26 & \textbf{47.10} & 1571.72 & \textbf{459.96} \\
m\_12\_t11 & 357 & 1387 & 70.73 & \textbf{56.95} & 1818.51 & \textbf{490.81} \\
m\_12\_t12 & 433 & 1708 & 355.25 & \textbf{209.85} & - & \textbf{3497.77} \\
m\_12\_t13 & 509 & 2046 & 908.61 & \textbf{704.83} & - & - \\
m\_12\_t14 & 583 & 2370 & 1450.02 & \textbf{1193.16} & - & - \\
m\_12\_t15 & 654 & 2675 & 1269.53 & \textbf{1070.46} & - & - \\
m\_12\_t16 & 713 & 2948 & 904.98 & \textbf{710.57} & - & - \\
m\_12\_t17 & 770 & 3200 & 518.96 & \textbf{412.98} & - & \textbf{5395.86} \\
m\_12\_t18 & 818 & 3408 & 272.96 & \textbf{215.98} & - & \textbf{2209.95} \\
m\_12\_t19 & 860 & 3593 & 125.81 & \textbf{102.35} & 5334.01 & \textbf{1019.15} \\
m\_12\_t20 & 895 & 3745 & 63.67 & \textbf{56.62} & 2388.44 & \textbf{527.68} \\
m\_13\_t11 & 391 & 1501 & \textbf{75.03} & 82.27 & 2167.30 & \textbf{857.24} \\
m\_13\_t12 & 468 & 1814 & 514.78 & \textbf{397.85} & - & \textbf{6130.18} \\
m\_13\_t13 & 550 & 2154 & 2347.92 & \textbf{1598.34} & - & - \\
m\_13\_t14 & 626 & 2487 & 8666.41 & \textbf{5328.73} & - & - \\
m\_13\_t15 & 704 & 2814 & - & \textbf{7710.58} & - & - \\
m\_13\_t16 & 775 & 3124 & - & \textbf{7171.84} & - & - \\
m\_13\_t17 & 842 & 3412 & 6989.75 & \textbf{4489.04} & - & - \\
m\_13\_t18 & 902 & 3670 & 4177.14 & \textbf{2747.70} & - & - \\
m\_13\_t19 & 958 & 3909 & 1752.16 & \textbf{1410.64} & - & - \\
m\_13\_t20 & 1003 & 4111 & 930.08 & \textbf{685.51} & - & \textbf{9567.04} \\
m\_13\_t21 & 1049 & 4296 & 475.52 & \textbf{324.19} & - & \textbf{4163.68} \\
m\_13\_t22 & 1085 & 4452 & 207.70 & \textbf{146.31} & 9481.00 & \textbf{1821.52} \\
m\_13\_t23 & 1116 & 4582 & 83.20 & \textbf{79.45} & 3989.72 & \textbf{756.90} \\
m\_13\_t24 & 1136 & 4673 & \textbf{36.93} & 37.21 & 1865.10 & \textbf{325.21} \\
\hline
\#Better & - & - & 4 & \textbf{56} & 0 & \textbf{29} \\
\#Solved & - & - & 56 & \textbf{60} & 15 & \textbf{29} \\
PAR-2 & - & - & 2841.65 & \textbf{1724.01} & 15989.31 & \textbf{12177.66} \\
\hline
\end{tabular}}
\end{table}


\begin{table}[!htbp]
  \centering
  \caption{SAT Competition 2022 Main Track Performance}
  \label{tab:sat2022-results}
  \setlength{\tabcolsep}{3.5pt} 
  \scalebox{0.73}{
  \begin{tabular}{lcccc}
    \toprule
    Solver & \#Solved & \#SAT & \#UNS & PAR-2 \\
    \midrule
    kissat (origin)   & 297 & 150 & 147 & 3116.98 \\
    kissat (ours)   & \textbf{301} & \textbf{151} & \textbf{150} & \textbf{2970.55} \\
    \midrule
    MiniSat (origin)  & 105 & 49  & 56  & 7963.79 \\
    MiniSat (ours)  & \textbf{127} & \textbf{58} & \textbf{69} & \textbf{7438.85} \\
    \bottomrule
  \end{tabular}
  }
\end{table}

Firstly, our methods significantly enhance the performance of CDCL solvers on complex arithmetic circuits.
Table~\ref{dataset1} summarizes the results of the first dataset. Each row reports the instance name, the number of variables and clauses, and the execution times of the competing solvers. 
It indicates that integrating our method reduces kissat’s average PAR-2 score by roughly 40\%, with improvements observed in more than 93\% of the instances. 
MiniSat also derives substantial benefits; after integration, it solves nearly twice as many instances as the baseline. In instances completed within the time limit, both before and after integration, our method helps improve MiniSat, achieving speedups of up to 5.74×, with an average acceleration of 3.88×. It outperforms the baseline in all these cases.

Secondly, our method is effective on general purpose benchmarks.
Table~\ref{tab:sat2022-results} summarizes the results on SC22.
The results show that our method improves both kissat and MiniSat on SC22, regardless of whether performance is measured by \#Solved, PAR2, or the counts of solved SAT and UNSAT instances (\#SAT and \#UNS). Specifically, the integration increases kissat’s total solved instances by 4 and reduces its average PAR-2 score by 4.7\%. 
MiniSat also shows substantial gains, with the total number of solved instances increasing by roughly 21\%. 
This experiment demonstrates that the method has good generalizability.

\subsection{Ablation Experiments}

\begin{table}[htbp]
  \centering
  \caption{Ablation Study on kissat and MiniSat}
  \label{tab:ablation-shared}
  \scalebox{0.73}{
  \begin{tabular}{lccccc}
    \toprule
     & \multicolumn{2}{c}{kissat} & \multicolumn{2}{c}{MiniSat} \\
    \cmidrule(r){2-3} \cmidrule(l){4-5}
    Configuration & \#Solved & PAR-2 & \#Solved & PAR-2 \\
    \midrule
    Stage 1 + Stage 2 & \textbf{60} & \textbf{1724.01} & \textbf{29} & \textbf{12177.66} \\
    Stage 1 only      & 56 & 2885.92 & 22 & 13647.31 \\
    Stage 2 only      & 53 & 4002.89 & 21 & 14123.44 \\
    \bottomrule
  \end{tabular}
  }
\end{table}



This subsection conducts ablation experiments to evaluate the separate contributions of stages 1 and 2 of our method, thereby confirming their impact on solver performance. Because our method primarily targets complex arithmetic circuit problems, all experiments here are conducted on the first dataset. Table~\ref{tab:ablation-shared} summarizes the results on kissat and MiniSat respectively.

The ablation study demonstrates that stages 1 and 2 must operate together; enabling either stage in isolation decreases the solved count and increases the PAR-2 score, whereas combining both stages recovers the full gains. This confirms that both the inherent lineage and the dynamic usage of learnt clauses are indispensable, and that ignoring either dimension in clause management leads to inefficiency. An effective scheme must therefore balance and exploit both aspects.

\subsection{Extensibility and Applicability Discussion}

\begin{figure}[t]
    \centering

    \begin{subfigure}[b]{0.49\columnwidth}
        \centering
        \includegraphics[width=0.9\columnwidth]{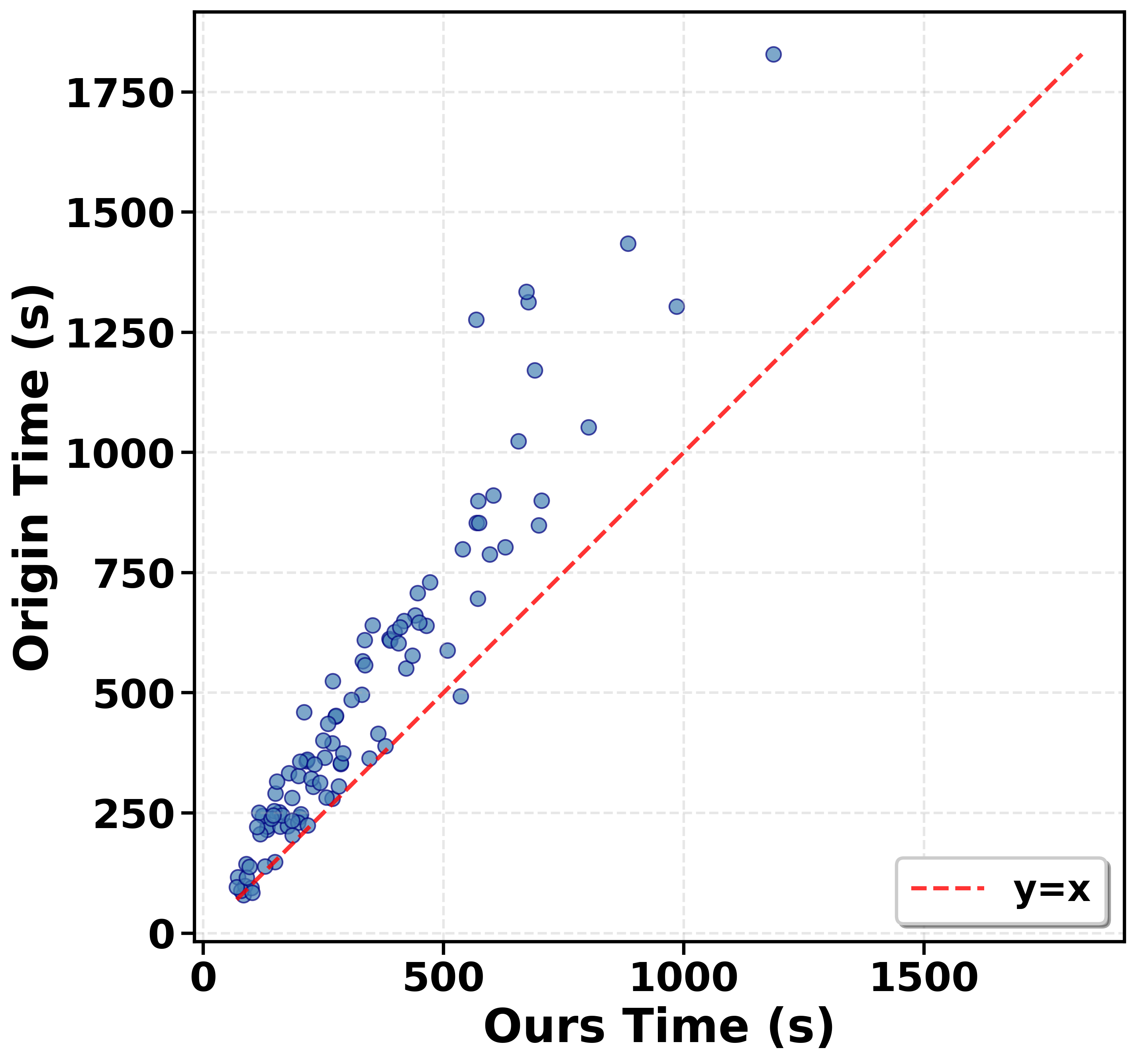}
        \caption{kissat}
        \label{fig:sub1}
    \end{subfigure}
    \hfill
    \begin{subfigure}[b]{0.49\columnwidth}
        \centering
        \includegraphics[width=0.9\columnwidth]{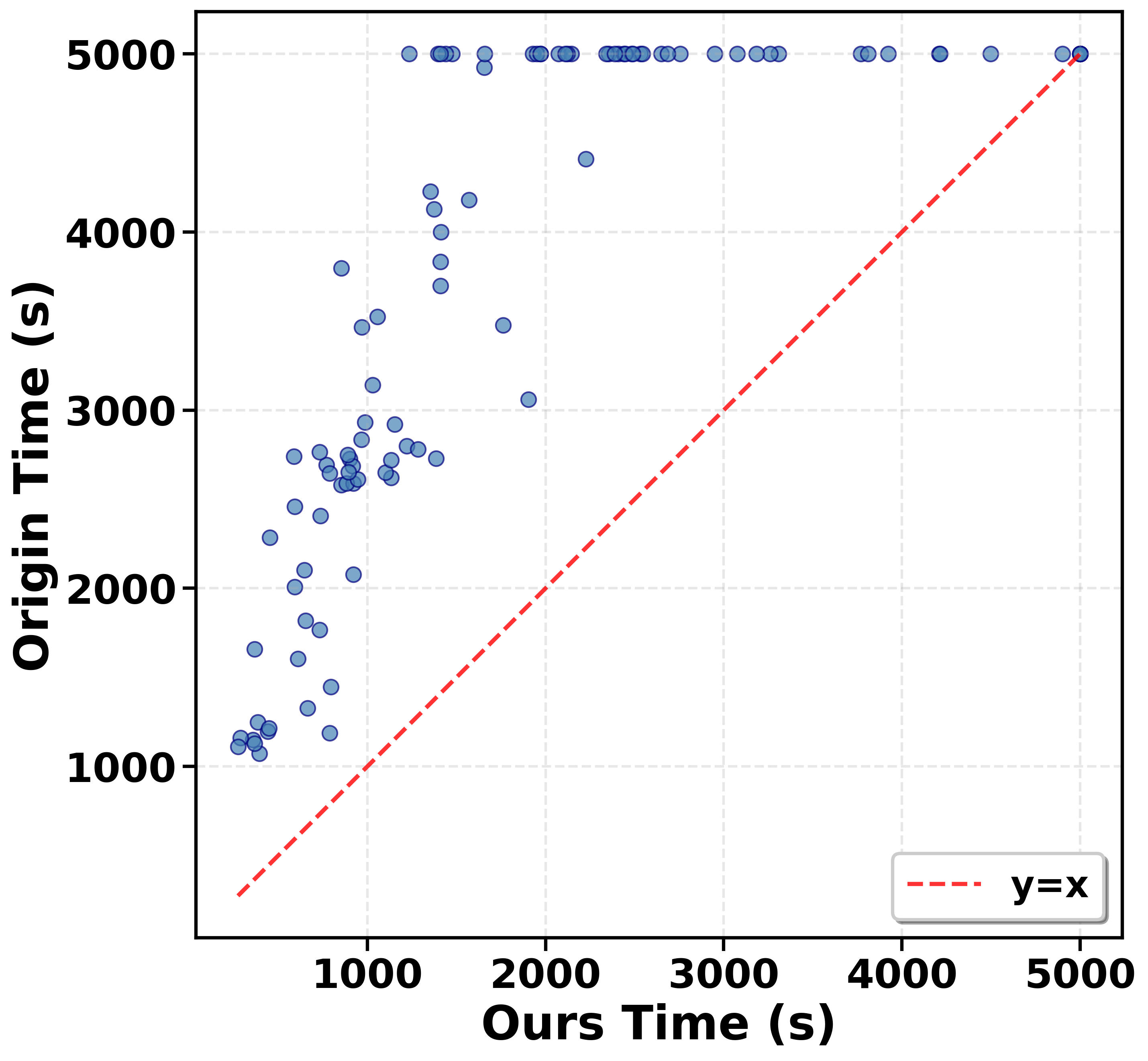}
        \caption{MiniSat}
        \label{fig:sub2}
    \end{subfigure}

    \caption{Comparison on Random 3-SAT Problems}
    \label{fig:R3SAT}
\end{figure}


\begin{table}[htbp]  
  \centering
  \caption{Runtime Comparison on Random 3 SAT Problems}
  \label{tab:Random3SAT}
  \scalebox{0.73}{
  \begin{tabular}{lccc}  
    \toprule
    Solver & \#Solved & PAR-2 & \#Better \\
    \midrule
    kissat (origin)   & \textbf{100}  & 490.51 & 5 \\
    kissat (ours)   & \textbf{100} & \textbf{331.71} & \textbf{95} \\
    \midrule
    MiniSat (origin)  & 56 & 5843.10 & \textbf{95} \\
    MiniSat (ours)  & \textbf{95} & \textbf{2044.65} & 0 \\
    \bottomrule
  \end{tabular}
  }
\end{table}


This subsection aims to further investigate the types of instances on which our clause management mechanism is effective.
The core motivation behind our approach is to address the loss of decision-level reasoning information in LBD-based clause quality evaluation when solving challenging EDA verification problems. To this end, we conduct experiments with cutoff 5000s.

\begin{table}[htbp]
\centering
\caption{Comparison on Cryptography Instances}
\label{Cryptography}
\scalebox{0.73}{
\begin{tabular}{lrr|rr|rr}
\hline
 &  &  & \multicolumn{2}{c|}{kissat} & \multicolumn{2}{c}{MiniSat} \\
Name & Vars & Clauses & origin & ours & origin & ours \\
\hline
grain55 & 2040 & 30236 & 380.2 & \textbf{270.02} & - & \textbf{4033.39} \\
bivium39 & 978 & 5514 & 1506.02 & \textbf{1129.7} & - & - \\
grain534 & 1854 & 27024 & 405.39 & \textbf{304.87} & - & \textbf{4711.79} \\
grain538 & 1849 & 27043 & 451.01 & \textbf{361.42} & - & \textbf{4944.61} \\
bivium-40 & 970 & 5432 & 565.18 & \textbf{424.37} & - & - \\
gus-md5 & 69561 & 226787 & - & \textbf{1951.39} & - & - \\
\hline
\#Better & - & - & 0 & \textbf{6} & 0 & \textbf{3} \\
\#Solved & - & - & 5 & \textbf{6} & 0 & \textbf{3} \\
PAR-2 & - & - & 2217.97 & \textbf{740.30} & 10000 & \textbf{7281.63} \\
\hline
\end{tabular}}
\end{table}

\paragraph{Evaluation on Cryptanalysis Problems}
This category shares key characteristics with multiplier verification instances, involving extensive XOR chains. Such problems exemplify a class of challenging real-world applications. The instances used in this subsection were collected from past SAT competition benchmarks, totaling 6 instances, all of which are unsatisfiable. 

Table~\ref{Cryptography} summarizes the comparison results for cryptography instances. Our findings indicate that the proposed method consistently enhances solver performance on this class of problems. With kissat, our approach surpasses the baseline across all instances and reduces the average PAR-2 score to approximately one-third of its original value. For MiniSat, the version augmented with our method successfully solves three instances, whereas the original solver fails to solve any.

\paragraph{Evaluation on Random 3-SAT Instances}
Random 3-SAT refers to random generated instances in which each clause of the CNF formula contains exactly three literals. 
These 100 unsatisfiable instances are generated using CNFgen~\cite{lauria2017cnfgen}, with variable counts ranging from 601 to 700, with the clause-to-variable ratio fixed at 7:1. 

The detailed comparision for each instance is given in the format of scatter plot in Figure~\ref{fig:R3SAT}.
Our method significantly improves both kissat and MiniSat. 
All 100 instances are solved by our kissat (PAR-2: 331.71), outperforming the original kissat in 95 cases (PAR-2 490.51), reduces the average PAR-2 score by 32.4\%, and achieves speedups of up to 2.25×. For MiniSat, 95 instances are solved (PAR-2: 2044.65), compared to only 56 with the original (PAR-2: 5843.10), boosting solved instances by approximately 70\%, reducing the PAR-2 score to under one-fifth of the original, and outperforming it on all non-timeout instances.

\paragraph{Discussion Summarize}
The above experiments consistently show that our method’s applicability extends far beyond complex arithmetic circuits, suggesting strong potential for broader applications in cryptanalysis and mathematics.
Moreover, our method’s performance advantage grows markedly on harder instances, underscoring its effectiveness in tackling challenging cases.

\section{Conclusion and Future Work}\label{Sec:Conclusion}

This paper challenges the long‑standing dominance of LBD for clause quality, particularly in complex arithmetic circuit verification. We decouple clause lineage from usage patterns and design a two‑stage, LBD‑free clause management mechanism. Experiments show substantial performance gains on arithmetic circuit benchmarks while maintaining state‑of‑the‑art results on general benchmarks.

This marks the beginning of approaches to challenge LBD‑based clause quality metric. Future work includes evaluating our method on broader problem classes, distinguishing clause roles in Stage 1 more precisely, and designing clause reduction triggers tailored to our decoupling approach.

\balance



\balance
\bibliographystyle{ACM-Reference-Format}
\bibliography{reference}

\end{document}